\documentclass[11pt,preprint]{aastex}

\shorttitle{Fragmentation in Sheets}
\shortauthors{Burkert \& Hartmann}

\begin{document}

\title{Collapse and Fragmentation in Finite Sheets}

\author{Andreas Burkert\altaffilmark{1} and Lee Hartmann\altaffilmark{2}}

\altaffiltext{1}{University Observatory Munich, Scheinerstrasse 1, D-81679 Munich, 
Germany}
\altaffiltext{2}{Harvard-Smithsonian
Center for Astrophysics, 60 Garden Street, Cambridge, MA 02138}

\email{burkert@usm.uni-muenchen.de, hartmann@cfa.harvard.edu}

\newcommand\msun{\rm M_{\odot}}
\newcommand\lsun{\rm L_{\odot}}
\newcommand\msunyr{\rm M_{\odot}\,yr^{-1}}
\newcommand\be{\begin{equation}}
\newcommand\en{\end{equation}}
\newcommand\cm{\rm cm}
\newcommand\kms{\rm{\, km \, s^{-1}}}
\newcommand\K{\rm K}
\newcommand\etal{{\rm et al}.\ }
\newcommand\sd{\partial}

\begin{abstract}
We present two-dimensional simulations of finite, self-gravitating gaseous
sheets.  Unlike the case of infinite sheets, such configurations do
not constitute equilibrium states but instead are subject to global
collapse unless countered by pressure forces or rotation.  The initial
effect of finite geometry is to promote concentrations of material
at the edges of the sheet.  If the sheet is not perfectly circular,
gravitational focussing results in enhanced concentrations of mass.
In the second-most simple geometry, that of an elliptical outer boundary,
the general result is collapse to a filamentary structure with the densest
concentrations of mass at the ends of the filament.  We suggest that these
simple calculations have interesting implications for the gravitational
evolution of overall molecular cloud structure, envisioning that 
such clouds might originate as roughly sheetlike sections of gas 
accumulated as a result of large-scale flows in the local interstellar medium.  
We show some examples of local clouds
with overall filamentary shape and denser concentrations of mass and star
clusters near the ends of the overall extended structure, suggestive of
our simple ellipse collapse calculations.  We suggest that cluster-forming
gas is often concentrated as a result of gravity acting on irregular boundaries;
this mechanism can result in very rapid infall of gas which may be of
importance to the formation of massive stars.  This picture suggests that
much of the supersonic ``turbulence'' observed in molecular clouds might
be gravitationally-generated.  Our results may provide impetus for further
theoretical explorations of global gravitational effects in molecular clouds
and their implications for generating the substructure needed for fragmentation
into stars and clusters.
\end{abstract}


\section{Introduction}

A central issue in star formation is the origin of the small-scale structure
in molecular clouds which leads to the creation of stars.  Many
researchers have suggested that this substructure is due to
``turbulence''; complex, often supersonic, motions lead to density
concentrations which then collapse to form stars (e.g., Larson 1992;
Elmegreen 1997; Mac Low \etal 1998; Padoan \& Nordlund 1999;
Klessen \& Burkert 2000, 2001; Ostriker, Gammie, \& Stone 1999;
Klessen, Heitsch, \& Mac Low 2000; Bate, Bonnell, \& Bromm 2002,
2003; Gammie \etal 2003; Li \etal 2004; see also review by
Elmegreen 2002).  However, the nature of these supersonic motions
is far from clear, making it difficult to evaluate the role
of turbulent fragmentation in star formation.  For example,
small-scale driving of turbulence is employed in many
numerical simulations to form stellar mass concentrations, but
this may not be consistent with large-scale structure
such as extended massive filaments seen in many clouds (e.g.,
Schneider \& Elmegreen 1979; Hartmann 2002).  Another unresolved
question is whether the periodic boundary conditions used in many
simulations can really capture the essential physics of real
clouds, in which material can either be accreted or ejected.
More broadly, the recent recognition that 
molecular clouds have short lifetimes (Elmegreen 2000; Hartmann, Ballesteros-Paredes,
\& Bergin 2001 $=$ HBB01) emphasize the likely role of initial conditions
in establishing the turbulent velocity field, an area which has not
been adequately explored.

A rather different approach to fragmentation was taken
by Larson (1985), who pointed out that infinite self-gravitating sheets and
filaments have a characteristic scale of fastest growth, typically
a few times the sheet or filament scale height.  In this scenario
of gravitational fragmentation, gravity
acts on a smooth distribution of material in a cloud of limited dimensionality
(sheet or filament geometry) but infinite extent to produce fragments
of finite mass.  This model
seems to avoid the need to put smaller (density or velocity) structure in ``by hand''.
Hartmann (2002) pointed out that the molecular cloud cores in Taurus
are elongated in the direction of their host filaments, in the sense
predicted by gravitational fragmentation.  However, the static initial
conditions assumed in the simplified Larson (1985) discussion are not
consistent with observed supersonic velocity dispersions (e.g., Mizuno \etal 1995). 
In addition, as we show below, discarding the assumption of infinite sheets
or filaments results in crucial modifications to Larson's picture;
finite sheets behave differently.

In this paper we extend Larson's investigations to consider structure formation
in sheets of finite sizes.  We focus on a simplified investigation of initially
homogeneous and isothermal sheets to isolate essential physics of the problem without
introducing complications due to heating and cooling processes or turbulent driving.
Even within this extremely limited set of conditions, we show that a rich variety
of fragmenting structures can arise in gravitationally collapsing finite sheets,
including multiple large concentrations that might lead to the
formation of clusters and massive stars, with
connecting and fragmenting filaments.  Our investigations suggest that
gravitationally-induced motions may be a significant and in many cases dominant
contributor to supersonic motions in molecular clouds.
We also suggest that boundary effects in general might play an important role;  
the assumption of periodic boundary conditions in simulations of cloud evolution
might therefore neglect important aspects. The importance of boundary effects has 
first been pointed out by Bastien (1983) who studied the collapse of cylindrical clouds.
Finally, we also speculate on
a possible way of relating turbulent structure and initial mass functions.

\section{Motivation}

HBB01 argued that molecular clouds
in the solar neighborhood are mostly formed as a result of large-scale flows,
which pile up atomic gas until sufficient column densities are accumulated to
shield the gas from the interstellar ultraviolet radiation field and allow molecules
to form.  The flows are presumed to be driven by stellar energy input, principally
supernovae.  The resulting clouds are then formed as wall sections of ``bubbles''
(e.g. Vazquez-Semadeni, Passot, \& Pouquet 1995;
Passot, Vasquez-Semadeni, \& Pouquet 1995; de Avillez \& Mac Low 2001;
Wada \& Norman 2001; HBB01, and references therein).

The simplest abstraction of this picture of cloud formation (which is also consistent with
cloud formation behind a shock front, such as in a spiral
density wave) is a flat uniform sheet of finite dimensions.
While real clouds formed by flows obviously will not initially be perfectly flat or have
uniform surface densities, it seems appropriate to make an initial exploration to isolate the effects of
gravity on sheets with finite structure.  To further simplify the analysis we assume
isothermality and consider sheets that are either static or have simple,
smooth velocity fields.  Even with these restrictive assumptions,
a wide variety of behavior results, which may have more general implications.

Before presenting the simulations, it is instructive to start by
considering some analytic results and approximations which illustrate some basic
properties.  We start with the simple
case of a static, isothermal, infinite, infinitely thin sheet with an initially constant
surface mass density $\Sigma_{\circ}$.  In this case the dispersion relation is
(Larson 1985)
\begin{equation}
\Gamma^2 ~=~ 2 \pi G \Sigma_{\circ} k ~-~ c_s^2 k^2\,, \label{eq:dispsheet}
\end{equation}
where $\Gamma$ is the exponential growth rate.
There is a critical wavenumber,
\be
k_c ~=~ 2 \pi G \Sigma_{\circ} / c_s^2\,, \label{eq:kc}
\en
above which no exponential growth is possible; i.e. there is
a minimum wavelength (a `Jeans' length) for gravitational instability.
Differentiating equation (\ref{eq:dispsheet}) with respect to $k$, one
can find the wavenumber at which the exponential growth is fastest,
\be
k_f ~=~ \pi G \Sigma_{\circ} / c_s^2 ~=~ k_c/2\,.
\en
This result suggests that the sheet will break up into fragments of
preferred mass
\be
M_f ~\sim~ \lambda_f^2 \Sigma ~=~ 4 c_s^4/G^2 \Sigma_{\circ}\,,
\en
where $\lambda_f = 2 \pi/k_f$.
Similar results hold for a self-gravitating sheet of finite thickness,
in hydrostatic equilibrium, with the critical wavenumber reduced by a factor of two.
Fragmentation of an infinite filament similarly occurs on some small multiple of the
thickness of the configuration (Larson 1985).

As pointed out by Larson (1985), although there is a critical wavelength for
gravitational collapse in a uniform density medium (the Jeans length), it is
difficult to fragment in such a situation, for instance in
a uniform density sphere, because the growth rate increases
monotonically with increasing wavelength (decreasing wavenumber); 
large-scale collapse tends to overwhelm fragments (Tohline 1980). 
In contrast to the uniform sphere case,
infinite sheet or filament models do exhibit a characteristic scale of growth.
However, these initial equilibrium
states require an infinite extent of the sheet or filament; and real clouds cannot
be infinite.  This leads to some important changes
in the sheet/filament picture of fragmentation.

The gravitational potential at a point $r$ from the center of an infinitely-thin,
uniform surface density sheet of radius $R$ is (Wyse \& Mayall 1942)
\be
\Phi = - 4 G \Sigma R \, E(r/R)\,, \label{eq:phisheet}
\en
where $\Sigma$ is the surface density and $E$ is the second complete elliptic integral.
The gravitational acceleration toward the center at $r$ is
\be
a_r = - {\sd \Phi \over \sd r} = 4 G \Sigma {R \over r} \left [ K(r/R) - E(r/R) \right ]\,, \label{eq:ar}
\en
where $K$ is the first complete elliptic integral.  The acceleration goes to infinity
at $d = R$, which would not occur in a sheet with finite thickness; thus we restrict
use of this equation to regions considerably more than a sheet thickness from the edge.

\begin{figure}
\epsscale{0.5}
\plotone{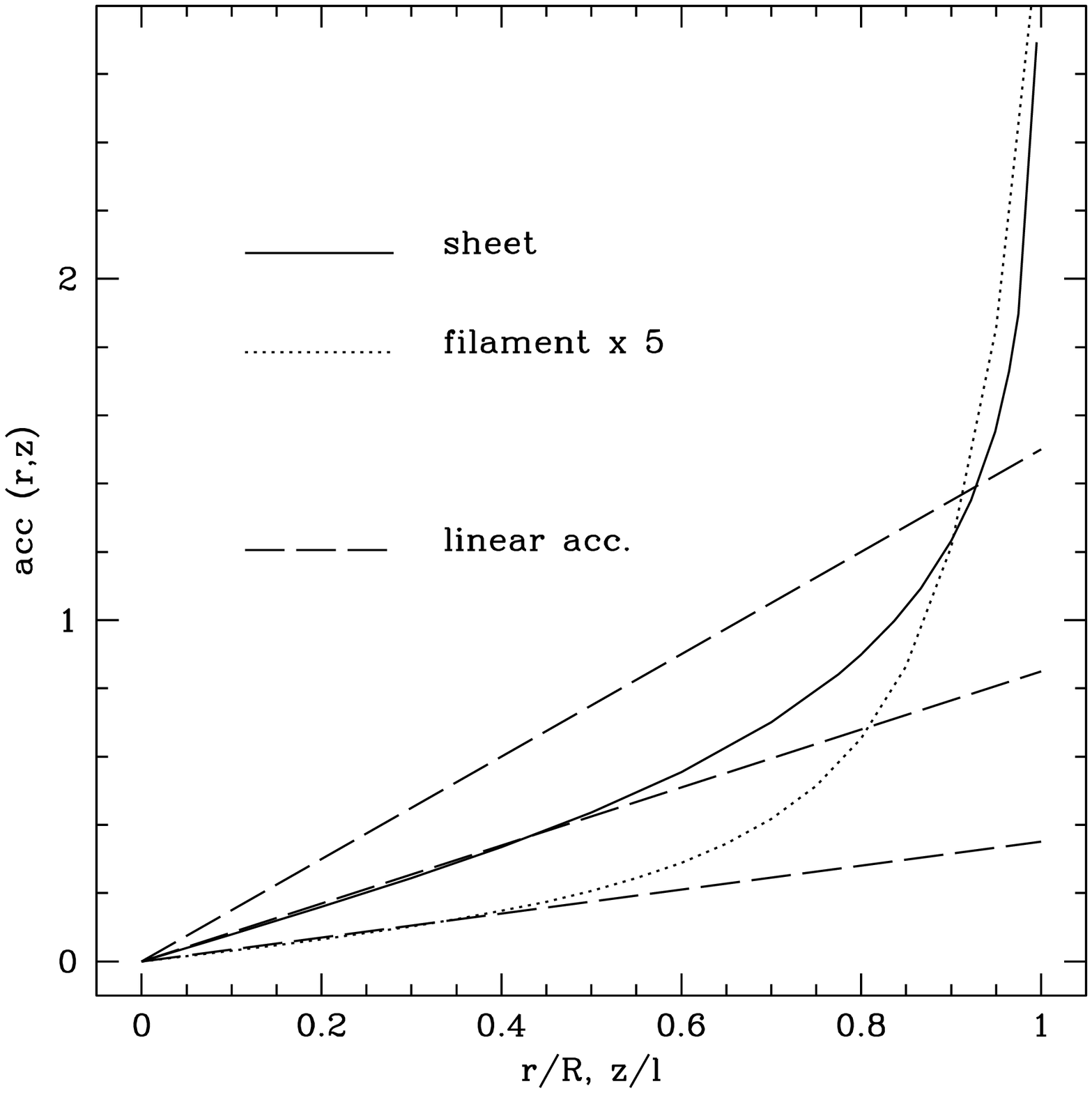}
\caption{Gravitational acceleration for the
uniform sheet (equation \ref{eq:ar}; solid line), in units of
$4 G \Sigma$, vs. radial distance in units of total sheet radius $r/R$,
along with the acceleration for the uniform filament (equation \ref{eq:filar};
dotted line), in units of $G \rho / 5$.  Various linear terms are indicated
by dashed lines (see text). }
\label{fig:linplot}
\end{figure}

Figure \ref{fig:linplot} shows the acceleration in units of $4 G \Sigma$ as a function of $r/R$.
The steep increase of inward acceleration as $r \rightarrow R$ implies that the sheet,
initially at rest, will immediately proceed to collapse, with material piling up
most rapidly at the outer edge (limited by gas pressure gradients which we ignore here,
i.e., we are assuming that the sheet contains many Jeans masses).

It is useful to estimate the timescale of global collapse for comparison with
numerical results.
Using the expansions of the $K$ and $E$ integrals (Abramowitz \& Stegun 1972),
equation (\ref{eq:ar}) can be written as
\be
a_r  = {1 \over 2} {d v^2 \over dr} =
\pi G \Sigma \left [ {r \over R}
+ {3 \over 8} \, \left ( {r \over R} \right )^3
+ {45 \over 192} \, \left ( {r \over R} \right )^5 + ...
\right ] \,. \label{eq:arexp}
\en
Ignoring pressure support, a collapse timescale $t_c$ can be estimated 
for a subregion of size $\delta v$ lying in the inner region of
the sheet of radius $R$.  Integrating equation 
(\ref{eq:arexp}) using only the linear term, starting from rest,
and assuming that $\Sigma$ does not change significantly within the inner
region (see \S 3.1), a typical infall velocity of the subregion is
\be
v^2 = {\pi G \Sigma \over R} (\delta r )^2\,, 
\label{eq:vregion}
\en
and thus
\be
t_c = {\delta r \over v}  = \left ( {R \over \pi G \Sigma} \right)^{1/2} \,.
\label{eq:sheettc}
\en
Note that $t_c$ is independent of the size of the region $\delta r$, a result
that will be used in the following section.
While this collapse timescale ignores the non-linear acceleration, and thus
does not describe the pile-up of material at the edge, we find numerically
that $t_c$ is a good estimate of the time it takes for the edge of the circular sheet to
fall to the center (\S 3.1).

Without rotation or some other motion,
the ultimate fate of this circular sheet is to collapse entirely to the center.
The dashed lines in Figure \ref{fig:linplot} show
linear forms for $a_r$; the middle dashed line indicates 
the situation where an outward acceleration is
comparable to the first term in the expansion of equation (\ref{eq:arexp})
balances the inward gravitational acceleration of the inner region.
Solid-body rotation, with centripetal acceleration $a(c)_r = - \Omega^2 r \propto r$,
where $\Omega = $~constant, could in principle be such
an example, preventing collapse in the inner sheet regions. 
However, the non-linear acceleration as $r \rightarrow R$
shows that such rotation cannot stop the edge from collapsing
to a ring whose dimensions are set by angular momentum.  Moreover,
the uniformly rotating sheet, whether in the non-equilibrium case of constant surface
density, or in the equilibrium case of $\Sigma \propto [ 1 - (r/R)^2 ]^{1/2}$
(Mestel 1963), is unstable to large-scale perturbations (Hunter 1963), and
generally results in large-scale redistribution of material with a concentration
of mass to the center (see, e.g., Binney \& Tremaine 1987, pp 374-375).
Conversely, large rotation (such as indicated by the upper dashed curve)
could prevent the inner region from collapsing,
but only at the expense of having the interior expand and the edge collapse
to an outer ring.

The finite filament exhibits similar behavior.
For a uniform cylindrical filament of radius $h$ and length $2l$.
the acceleration toward the center at a point on axis lying a distance $z$ from the
center is
\be
a_z = - 2 \pi G \rho \left [ 2 z - (h^2 + (l + z)^2 )^{1/2}
+ (h^2 + (l - z)^2 )^{1/2} \right ] \,, \label{eq:filar}
\en
where $\rho$ is the density of the filament.  When
$l \gg h$ and considering points away from the exact end of the filament, $l-z \gg h$,
\be
a_z \approx - \pi G \rho h^2 \left [(l+z)^{-1} + (l-z)^{-1} \right ]\,.
\en
But $\pi \rho h^2 = m$, the mass per unit length of the filament.  Using this relation,
and expanding the quantity in brackets, we have
\be
a_z \approx - G m \left [ {2 z \over (l^2 - z^2)} \right ]\,.
\en
Figure \ref{fig:linplot} also shows the acceleration of a thin filament in units of $G \rho$.
Note that the acceleration of an infinitely thin filament goes to
infinity at its edge (equation (11)), just as in the case of the infinitely
thin sheet, but this singularity disappears for finite $h$ (equation (10)).

As in the case of the sheet, solid-body rotation of the filament
(lower dashed curve) can help stabilize the
collapse of the inner regions, but cannot prevent the ends of the filament from
collapsing initially.  Alternatively, if one wants to prevent the filament ends of
collapsing, the solid body rotation would force the inner regions to expand
away towards the ends of the filament, resulting in concentrations at the endpoints.

These simple considerations illustrate the universal tendency for material to
pile up and concentrate at edges of finite structures due to gravity.
Whether the local sheet fragmentation can take place as envisaged by Larson (1985)
depends upon whether the {\em global} collapse overtakes or prevents {\em local}
collapse.  This is investigated numerically in the following section.

\section{Numerical simulations}

The numerical calculations are performed on a two-dimensional Eulerian, Cartesian
grid. The full computational region with dimension $2 \times L$ is represented
by a grid, composed of $N \times N$ grid cells, equally spaced in both directions.
Under the assumption of isothermality, the relevant differential equations to be
integrated are the  hydrodynamical continuity and momentum equations:

\begin{eqnarray}
\frac{\partial \Sigma}{\partial t} + \vec{\nabla} \cdot (\Sigma \vec{v}) & = & 0 \\
\frac{\partial \vec{v}}{\partial t} + (\vec{v} \cdot \vec{\nabla}) \vec{v} & = &
-\frac{\vec{\nabla} P}{\Sigma}- \vec{\nabla} \Phi \nonumber
\end{eqnarray}

\noindent where $\Sigma (\vec{x})$, $P(\vec{x})$ and $\vec{v}(\vec{x})$ are the
gas surface density, pressure and two-dimensional velocity vector at position
$\vec{x}$, respectively. The gravitational potential $\Phi$ is determined, solving Poisson's
equation in the equatorial plane (Binney \& Tremaine 1987)

\begin{equation}
\nabla^2 \Phi = 4 \pi G \Sigma
\end{equation}

\noindent with G the gravitational constant. The isothermal equation of state

\begin{equation}
P = c_s^2 \Sigma
\end{equation}

\noindent determines the pressure for a given surface density $\Sigma$ and sound speed $c_s$.

This set of equations is integrated numerically by means of an explicit finite
second-order van Leer difference scheme including operator splitting and
monotonic transport as tested and described in details in Burkert \& Bodenheimer (1993,1996).
In order to suppress numerical instabilities, an
artificial viscosity of the type described by Colella \& Woodward
(1984) is added (Burkert et al. 1997).

The Poisson equation is integrated on
the grid under the assumption that there is no matter outside of the computational region.
As we are focussing here on the gravitationally unstable sheets that collapse
towards the center of the region, outflow of gas beyond the outer boundaries can
be neglected. Therefore the outflow velocities at the outer boundary
are set to zero and a negligible pressure gradient is assumed. Most calculations were
typically performed with $100^2$ grid cells of size $\Delta = 2L/N$ and height
$\Delta$, where $2L$ is the largest dimension of the rectangular computational region.
Test calculations with N=200 and N=500 did not result in significant differences.
In these calculations the code units were set such that $G = 1$.

\subsection{Static circular sheet}

\begin{figure}
\epsscale{1.0}
\plotone{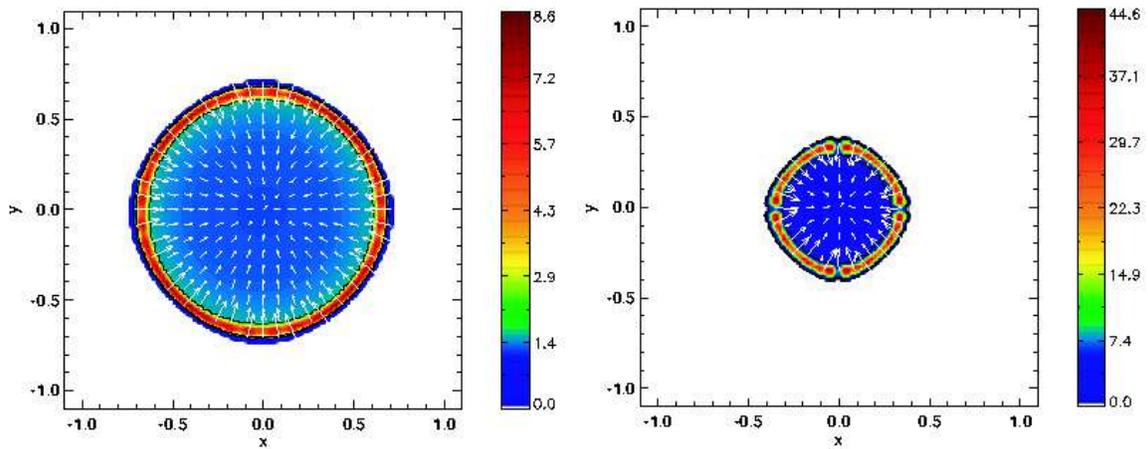}
\caption{Numerical simulation of the collapse of the finite two-dimensional
sheet.  In this calculation the initial sheet surface density is set to unity,
$G = 1$, and the sound speed is 0.1 in appropriate units (see text).
In code units, the time at which the snapshot is taken in the left panel
is 0.285; for the right panel, $t = 0.429$. The gaps in the outer ring on
the x and y axis in the left panel are purely numerical and caused by the representation
of the circular sheet by a cartesian grid.}
\label{fig:circle}
\end{figure}

Figure \ref{fig:circle}
shows the evolution of a static sheet with initially uniform surface
density (in code units, $\Sigma = 1$) of circular shape, with $R = 1$, and  
sound speed $c_s = 0.1$ 
(inside a computational region of $L = 1.1$).  
The mass of this sheet is thus $\pi \Sigma R^2 = \pi$
in code units.  Larson (1985) notes that the Jeans mass
for circular modes in an infinite thin static sheet is
\be
M_c = 1.17 c_s^4/(G^2 \Sigma)\,;
\en
thus this sheet initially contains $\sim 10^4$ Jeans masses. 
As expected, material initially piles up at the edge (left panel).
Note that even at an early stage collapse in the inner regions
is noticeable.  The evolution of the sheet is simple; the edge
grows as it falls in, and the entire structure collapses (right panel).

We never found any evidence for gravitational growth of 
fragments in the inner region, even for $c_s = 0$.  
Some fragmentation is seen in the piled-up ring material, which is
due to growth of initial numerical noise especially on the x and y-axes,
much of which is generated by the initial structure of a circular edge 
approximated in a rectangular grid.
More and earlier fragmentation in the ring occurs as the sound speed is
decreased.  The details of fragmentation in this and the further simulations
to be discussed should not be believed, as resolution quickly becomes
an issue; here we are concentrating on global structure.

We ran a number of simulations for
differing values of the sound speed; as long as the initial sheet contained
many Jeans masses, i.e., the sheet was sufficiently cold, the results were similar. 
For warm sheets of a few Jeans masses, fragmentation due to numerical fluctuations
in the edge ring was suppressed.  Finally, if the mass of the sheet was
small, the sheet ``bounced'' and then eventually adjusted to a static equilibrium.

\begin{figure}
\epsscale{0.8}
\plotone{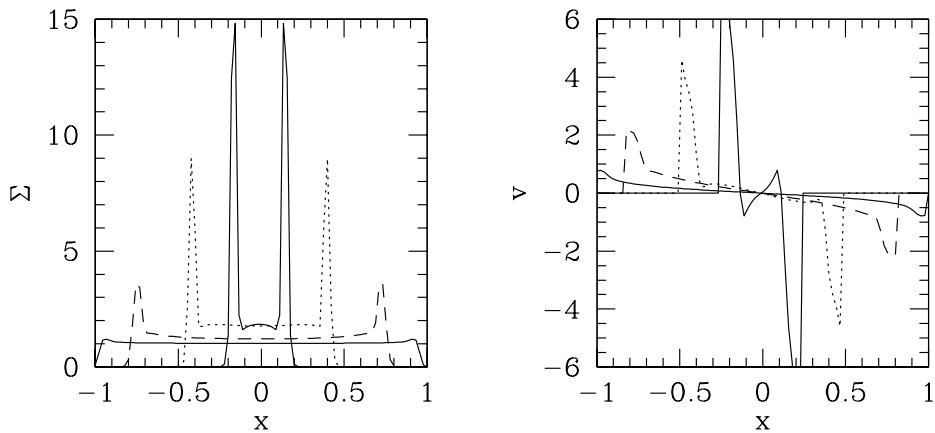}
\caption{Evolution of the density and velocity of the circular sheet collapse of
Figure \ref{fig:circle}.  The snapshots are taken at times $t =$ 0.09, 0.25, 0.41, 
and 0.49.  Note the flat density distribution and the linear velocity 
gradient in the inner regions, as expected from the result in equation
(\ref{eq:vregion}), until late in the evolution, when inner material 
begins to fall outward due to the gravity of the infalling edge.} 
\label{fig:circlerhov}
\epsscale{1.0}
\end{figure}

Figure \ref{fig:circlerhov} shows the density and 
velocity structure of the simulation shown in Figure \ref{fig:circle}.
Note the pile-up of material, and 
also that infall develops rapidly in the inner regions as well, as expected
from the analytic results of the previous section.
The collapse timescale in the linear (inner sheet) regime,
equation (\ref{eq:sheettc}), is $\pi^{-1/2} = 0.564$ in code units.
For the particular case described above, the time taken for the edge
to reach the center is approximately $t_g \sim 0.51$, i.e., slightly
shorter than the linear timescale.  

The above result for the timescale of global vs. local collapse helps to explain
why we never found any indication of
small-scale, linear perturbations becoming large before the entire sheet collapsed.
The infall which develops rapidly in all
sheet regions apparently invalidates the linear analysis of the infinite, static sheet.
Consider the following argument.  The most favorable location for a finite perturbation
(larger than a Jeans length) to grow before being overtaken by the general
collapse is at the center of the sheet, where the edge material
takes the longest time to arrive.
We may use the result of equation (\ref{eq:sheettc})
to evaluate the timescale of local collapse in the limit of zero sound speed
(negligible gas pressure) because this term simply represents gravitational
acceleration. Moreover, as shown in Figure \ref{fig:circlerhov}, even during the collapse
the surface density tends to remain uniform and the velocity gradient similar
until the infalling ``edge'' material overtakes it.  Now, equation
(\ref{eq:sheettc}) indicates that the timescale for collapse is
{\em independent} of the radius of the perturbed region; moreover, we find 
numerically that the time for the edge material to reach the origin is slightly
less than this value.  Thus, small perturbations cannot amplify before being swept up
by the overall collapse.  Because the collapse time (\ref{eq:sheettc}) is
proportional to $\Sigma^{-1/2}$, only very non-linear perturbations have
a chance to grow before being swallowed up by the global collapse.  
The situation is analogous to the collapse of a uniform
sphere, for which all radii reach the center at the same time, preventing
effective fragmentation from small perturbations (Larson 1985).

\begin{figure}
\epsscale{0.7}
\plotone{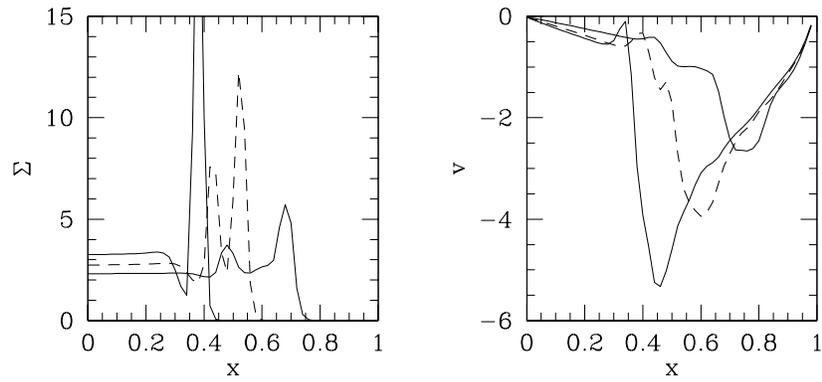}
\caption{Evolution of the uniform circular sheet with a ring-like
perturbation of excess surface density at times $t =$ 0.15, 0.21 and 0.26.
The perturbation grows as the sheet collapses, but not non-linearly, and is eventually
overtaken by the collapse of the edge (see text)}
\label{fig:circlep}
\epsscale{1.0}
\end{figure}

Even fairly large perturbations have difficulty growing before overall collapse
of the edge wins.  This is shown in Figure \ref{fig:circlep}, where we show
the evolution of an initial 10\% ring-like perturbation as a function of time.
The surface density of the perturbation grows linearly with time but never
outruns the edge, the latter eventually overtaking it. 

\subsection{Static ellipse}

\begin{figure}
\centering
\plotone{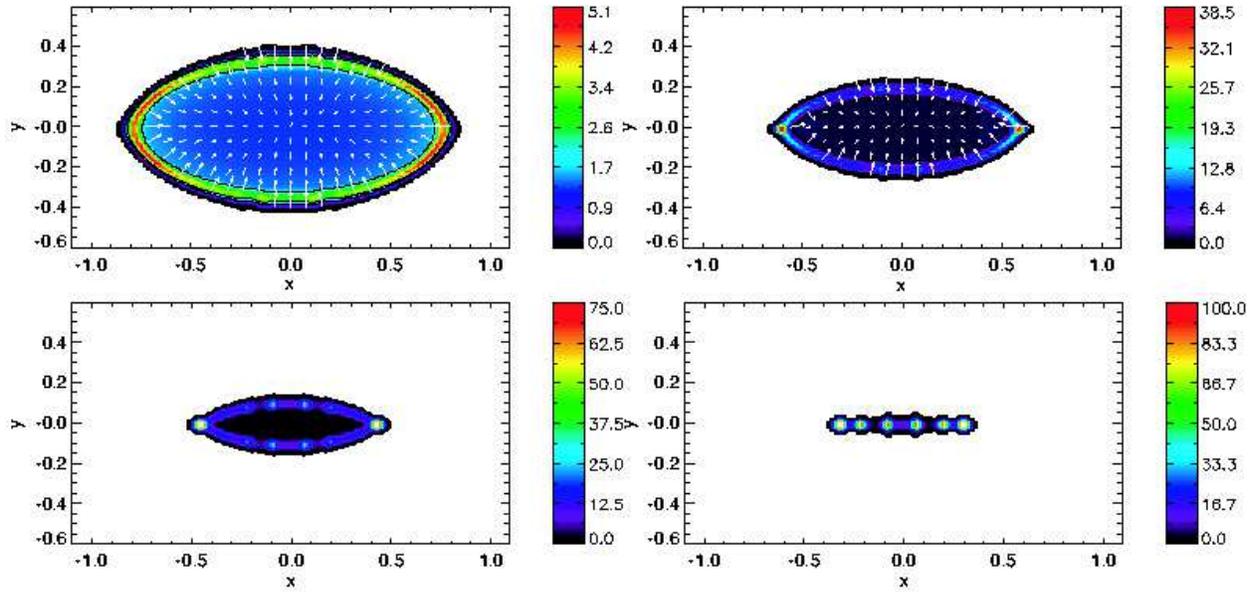}
\caption{Collapse of a static elliptical sheet.  Upper left: 
$t = 0.23$, material piles up at the edge as the sheet collapses.  
Upper right: $t = 0.33$, material accumulates particularly at focal
points at the ends of the ellipse.  Lower left: $t = 0.39$,
focal points become more prominent.  Lower right: $t = 0.44$, collapse
to a filament has occurred, with major concentrations at the ends.
}
\label{fig:ellipse}
\end{figure}

Figure \ref{fig:ellipse} a-d shows the evolution of an elliptical sheet with
an initial ellipticity of $e = 0.6$.  Again, we assume
$c_s = 0.1$.  As in the case of the circular sheet, material piles up at
the edge as the entire configuration collapses.  However, a new feature
arises: specifically, ``focal points'' appear, where gravity acts on the curvature
of the sheet edge to produce large, dense mass concentrations close to or outside
of the foci of the initial elliptical structure 
(upper right panel).  These mass concentrations grow with respect to the
rest of the edge material by gravitationally attracting neighboring material
to fall into them (lower left panel).  Finally, the sheet collapses into a filamentary
structure with large mass concentrations at both ends.

\begin{figure}
\epsscale{0.25}
\plotone{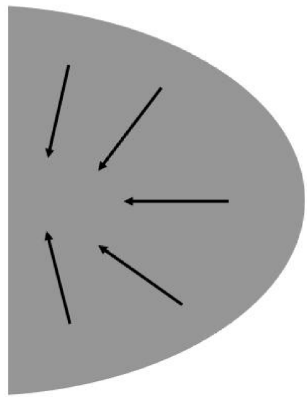}
\caption{Schematic geometry leading to mass concentration
at a ``focal point'' (see text)}
\epsscale{1.0}
\label{fig:focus}
\end{figure}

The geometry leading to focal points is indicated schematically in Figure (\ref{fig:focus}).
Any sheet edge which locally has a smaller radius of curvature than the larger-scale sheet
geometry will yield a local focus for gravitationally-collapsing material.

As before, our limited resolution prohibits any quantitative analysis 
of the number, mass, and sizes of fragments which eventually condense along the filament.  
The inhomogeneities present along
the filament are the result of numerical noise and limited resolution (e.g., Truelove \etal
1997) which get amplified during collapse.
Our main points are simply that the elongated sheet not only tends to collapse to a filament,
and that focal points develop which result in larger concentrations of mass at the filament
ends, a result seen, for example, in simulations of the collapse of elongated gas clouds
(Bastien 1983; Bonnell \etal 1991; Burkert \& Bodenheimer 1993).  

\subsection{Sheets without sharp edges}

\begin{figure}
\centering
\plotone{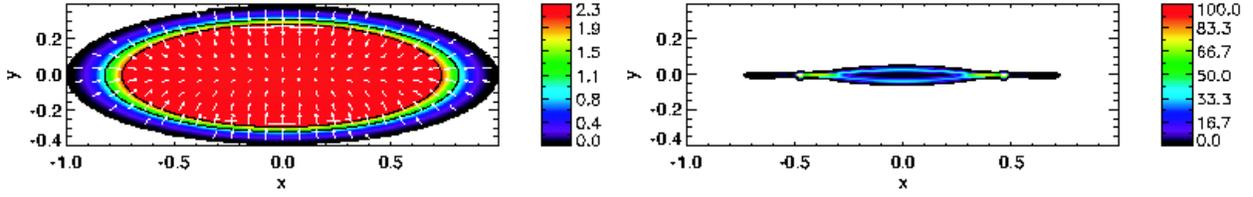}
\caption{Evolution of the static elliptical sheet with density decreasing to the edge,
The main features of the uniform ellipse are retained; a pile-up of material still occurs,
though on a smaller scale, more material lags outside the edge (left panel), and the final
filament formed shows mass concentrations somewhat interior to the ends of the filamentary
gas (right panel) (see text)}
\label{fig:nosharp}
\end{figure}

The previous calculations assumed a sharp outer edge, where the surface density decreases by
two orders of magnitude.  It seems implausible that real sheet-like clouds should have such
sharp edges, so we investigate a case in which the transition at the cloud boundary is more
gradual. 
Figure \ref{fig:nosharp} shows what happens when the density distribution of the ellipical sheet
falls off toward the edge.  In this particular case the density is made to fall off smoothly
to zero starting at 80\% of the distance to the elliptical boundary.  As shown in the left
panel, an edge concentration still develops, but in a smaller structure;
there is a modest amount of material outside this edge.  Focal points develop as before.
Finally, as shown in the right panel, collapse to a filament once again occurs, with mass concentrations
near the end, but now lower-density material extends outside of the focal point concentrations.
Note that Nelson \& Papaloizou (1993) found that spheroids did not necessarily form
concentrations at each end if the density distribution tapers off sufficiently. 
Uniform-density spheroids tend to have larger masses per unit lengths at their 
centers than the corresponding uniform sheets, suggesting that the difference
between two and three dimensions can be important.

\subsection{Expanding sheets}

\begin{figure}
\centering
\plotone{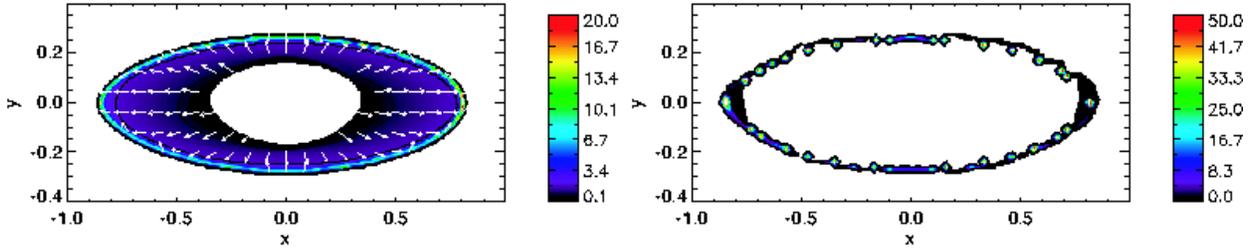}
\caption{Results for an expanding elliptical sheet.  Material moves outwards from the
origin to add to the edge, which still forms a concentration (left panel); ultimately,
most of the mass ends up in the expanding edge, with fragments determined by numerical
noise and resolution (right panel) (see text)}
\label{fig:expand}
\end{figure}

If sheets are made as parts of the walls of ``bubbles'' driven by supernova explosions
or stellar winds, they will generally exhibit some expansion in the direction perpendicular
to the main flow.  In our two-dimensional approximation, ignoring the bubble wall curvature,
we can introduce a similar effect by putting in a linear expansion term. 
Figure \ref{fig:expand} shows an expanding case, which was designed such that gravity was
not strong enough to reverse the expansion in the inner region, but still large enough
to play a role at the outer edge.  The initial radial velocity was assumed
to increase linearly with distance from the center and the surface density was constant.
Note that 
the entire region expands, but there is still a pile up of material at the edge and the
formation of focal points.  Thus expansion does not qualitatively change the mass concentration,
though it prevents the overall collapse of the sheet.

\subsection{Rotating sheets}

\begin{figure}
\centering
\plotone{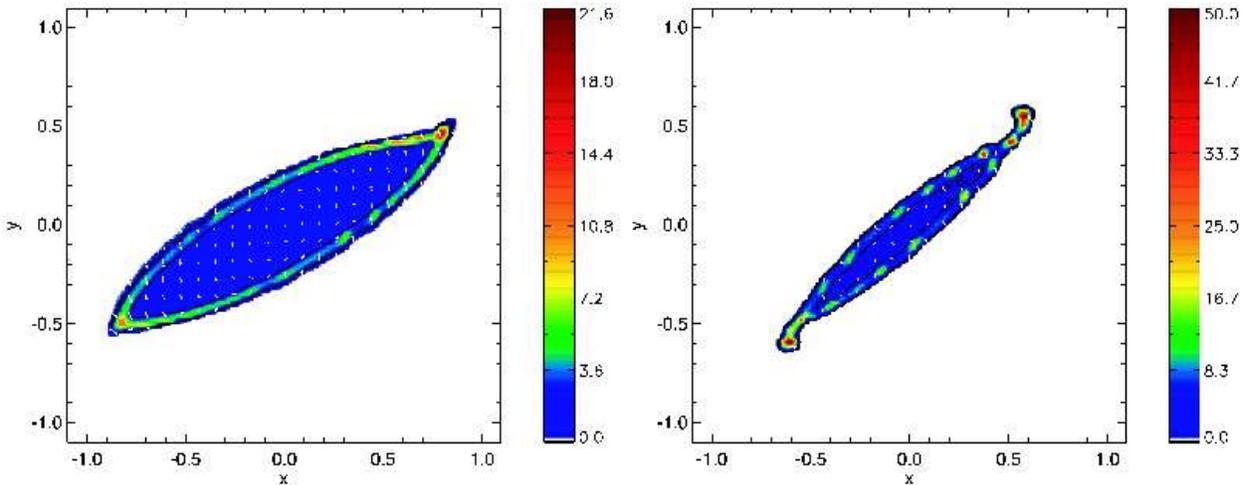}
\caption{Collapse of the rigidly rotating elliptical uniform sheet.  Focal points form and an eventual
filament results (see also following figure) which has significant rotational support against
gravity (see text)}
\label{fig:rotateellipse}
\end{figure}

\begin{figure}
\centering
\plotone{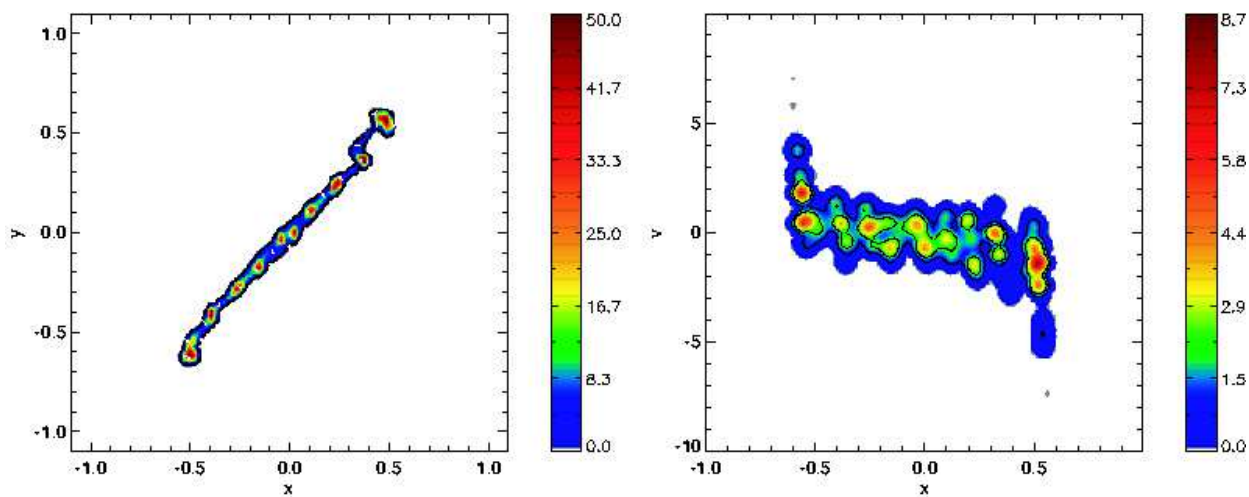}
\caption{Left panel: formation of a filament from the rotating ellipse simulation
(Figure \ref{fig:rotateellipse}).  The size and number of subfragments are not quantitatively
reliable.  Right panel: contours of constant surface density integrated in the y-direction
as a function of velocity in the x-direction (see text)}
\label{fig:rotateellipse2}
\end{figure}

In general there can be some angular momentum present in the plane of the sheet.  
Figure \ref{fig:rotateellipse} shows the case of an uniformly rotating, elliptical sheet.  Again
focal points form and collapse to a filament eventually occurs, with larger mass concentrations
at the ends of the filament (see also earlier work by Bonnell \etal 1991, 
Nelson \& Papaloizou 1993, and Monaghan 1994.) The rotation of the resulting filament 
(Figure \ref{fig:rotateellipse2}, left panel) is sufficient to slow the overall
collapse.  Material along the filament starts to be pulled in by the focal point
concentrations near the ends of the filament. 

We again emphasize that the number and properties of fragments in the
filament is not quantitatively reliable. 
(Note that fragmentation of rotating filaments
has also been found e.g. by Monaghan (1994).)
Nevertheless, as a qualitative
result it is interesting to inquire what velocity structure would be 
seen by an observer
in the plane of the sheet (now the plane of the filament).  
The right panel of Figure \ref{fig:rotateellipse2} shows 
a contour plot of surface density
integrated along the y-direction as a function of the velocity in the x-direction,
when the two edges have merged into the connecting filament. 
In addition to the evident overall rotation, there are local velocity perturbations
due to the gravitational accelerations of the various mass concentrations,
(an effect earlier seen in the simulations of fragmenting cylindrical clouds by  
Bonnell \& Bastien 1993; e.g., their Figures 3-5).
The original total mass of the ellipse was 1.26 in code units, 
and the initial major axis was
unity, so one would expect gravitationally-induced velocities to be of order
$v_{dyn} \sim (GM/R)^{1/2} \approx$~unity; the overall velocity gradient in the
line of sight is in agreement with this estimate.  The ``turbulence'' in the line
of sight, i.e., local fluctuations due to gravitational perturbations by local
concentrations, is also of this order.  The qualitative idea that differing
mass concentrations along a filament can (and must) induce smaller-scale 
velocity structure (``turbulence'') is worth noting.  
Our results bear an interesting qualitative resemblance to the velocity gradients
seen in the $^{13}$CO emission of the Orion A cloud (Bally \etal 1987; see \S 4.2).

\subsection{Ellipse with surface density gradient}

\begin{figure}
\centering
\plotone{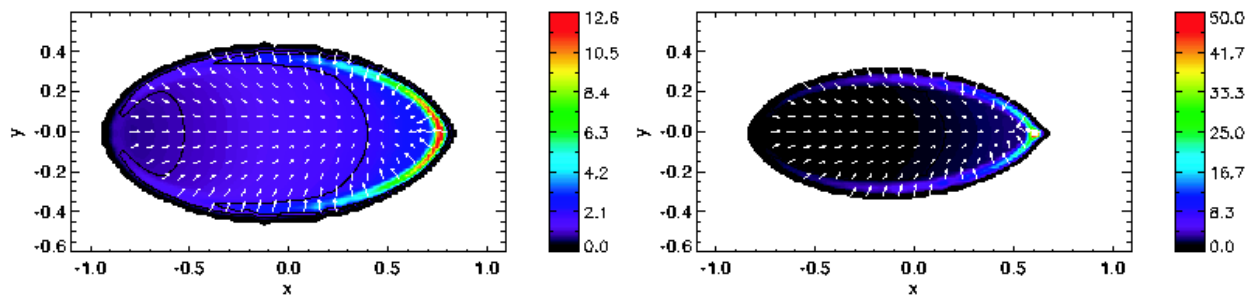}
\caption{Evolution of the static elliptical sheet with a linear surface density gradient along
the major axis.  The resulting evolution is similar to the uniform ellipse, except that the
dense edge and focal point concentration develop only at one end (see text)}
\label{fig:siggrad}
\end{figure}

It would be surprising if sheetlike clouds in the interstellar medium were uniform in surface
density.  We consider the next most complicated case, that of a uniform linear surface density
gradient along the major axis of the elliptical sheet.  Figure (\ref{fig:siggrad}) shows what
happens in the case where the surface density varies by a factor of four from one end of the
ellipse to the other, with $\Sigma = 2$ at the right end and $\Sigma = 0.5$ at the left.
The evolution is basically the same as that of the uniform ellipse, except that the dense
edge and focal point develop only at the dense end.

\subsection{``Ghosts''}

\begin{figure}
\centering
\plotone{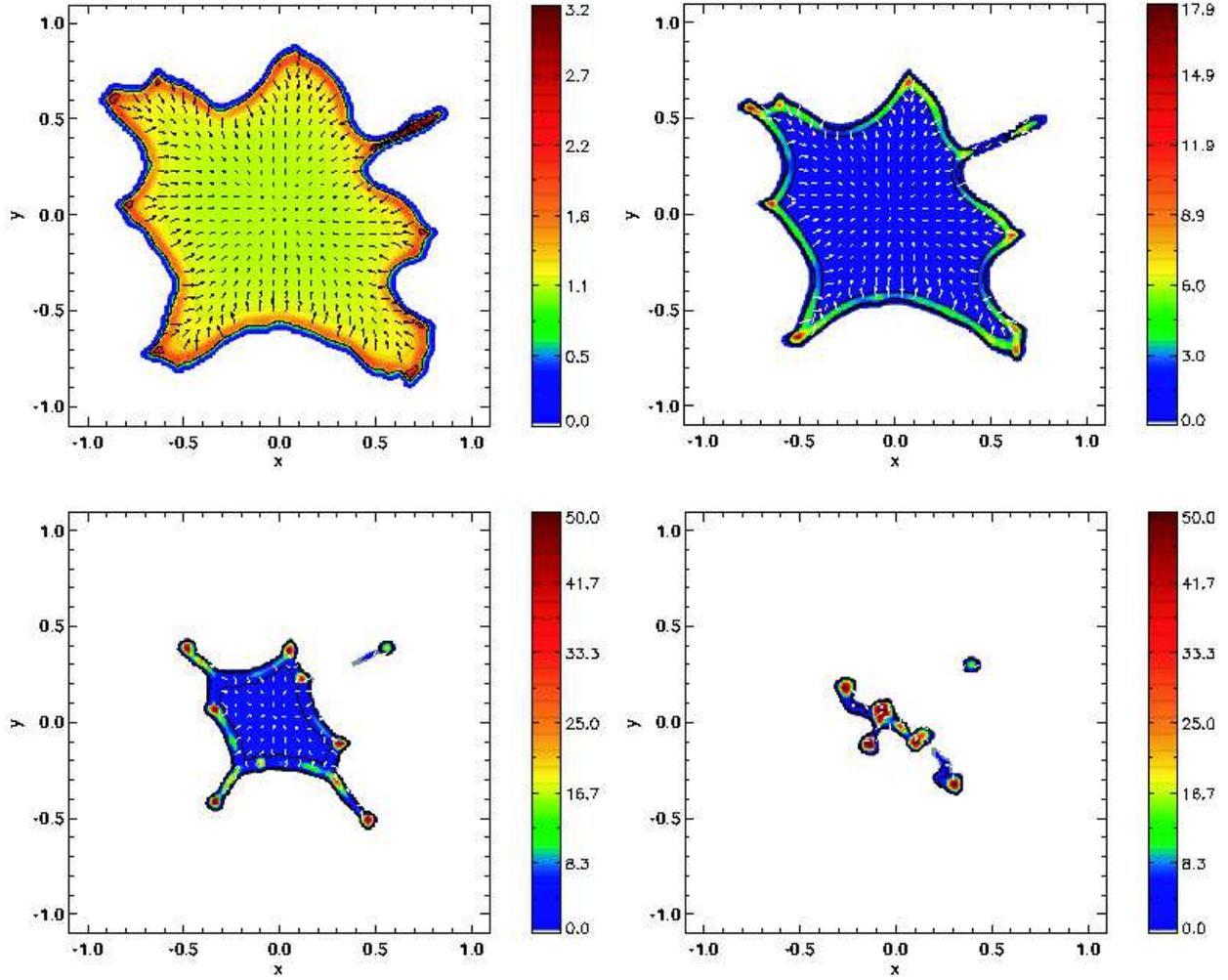}
\caption{Collapse of the ``ghost'', a sheet with highly
irregular boundary  (see text)}
\label{fig:ghost}
\end{figure}

It would be surprising if real clouds had perfectly smooth boundaries of either circular or
elliptical shape.  As indicated schematically in Figure (\ref{fig:focus}), any irregularity
with a small radius of curvature will tend to produce a concentration.  To explore the qualitative
nature of a complex boundary, in Figure \ref{fig:ghost} we show the results of the
collapse of a sheet with uniform initial surface density but an arbitrary irregular boundary 
(the ``ghost'').  As shown in the sequence of figures, pile up of material occurs first
along the edge, as before; focal points develop soon after.
As collapse proceeds, more material is pulled into the focal concentrations, which
fall in toward the origin. Near the end of the calculation, most of the mass lies in
concentrations, in number initially reflecting the number of initial ``nodes'' in the original
boundary; merging and subsequent evolution probably occurs but we cannot follow it in detail
with our resolution.

\section{Discussion}

\subsection{Initial conditions}

Our results show the powerful tendency of finite self-gravitating sheets to develop structure as a result
of gravitational focussing.  This immediately raises the question: how relevant are these highly simplified
calculations?  Real clouds are likely to have much more initial structure than what we have imposed in
our simulations; however, this should simply generate more substructure due to local focussing effects.
Similarly, the overall tendency for a non-circular sheet to collapse to a filament should also
be robust; more initial substructure will not stop the global collapse to a filament,
though the detailed structure could be much more complex.

Because a finite self-gravitating sheet will immediately start to collapse at its edges,
our assumption of static initial edge structure is probably not very realistic.  However, we think that
this complication does not matter very much.  As real molecular clouds are accumulated out of material
in the diffuse interstellar medium, collapse will start, leading to 
concentrations at edges some time before the final cloud mass has been accreted; 
but because this process is so rapid, it is not
important whether or not this is regarded as an initial condition or as an early development.
Perhaps our results for the sheet with a decline in density near 
the outer edge (Figure \ref{fig:siggrad}) can be thought of as indicating the evolution in a case
where material is still being accumulated in outer regions as the interior collapses.
As discussed in the previous sections, such edge effects can only 
be avoided by using substantial differential
rotation or internal pressure gradients
in ways that are not clearly relevant to most molecular clouds.

Of course, the formation of real clouds by flows will introduce some density inhomogeneities
and velocity perturbations.  Thus one can expect the structure of real
clouds to develop in a much more complex way than considered here.  But we suggest,
as demonsttrated in the following subsection, that our results may be relevant to the large-scale
or overall morphology of at least some molecular clouds, with significant substructure superimposed
by velocity and density perturbations.

Broadly speaking, our results are a simple case of the more general proposition of
Ballesteros-Paredes, Vazquez-Semadeni, \& Scalo (1999) that molecular clouds cannot
be in virial equilibrium.  As a technical matter, our calculations also suggest
that the occasional practice in numerical simulations of ``turning on'' gravity
after some evolution is not appropriate; gravity has long-range effects which
must be considered.  In addition, it seems clear that computational boxes with
periodic boundary conditions will not capture potentially important evolution. 

\subsection{Cloud morphologies}

It is obvious that a huge variety of shapes and fragments can result from
sufficiently complicated initial conditions at sheet edges, or from
a spectrum of density fluctuations within sheets.  Taking the larger view,
it is interesting to note that already the second-simplest figure - an elliptical sheet -
produces filaments with larger mass concentrations at each end.  This result suggests that
as clouds initially are likely to be non-circular even if sheetlike
configurations of this type might be fairly common.
Here we briefly consider the morphology of some well-known local star-forming regions
in light of our simplified collapse calculations.

\begin{figure}
\includegraphics[width=4in,angle=-90]{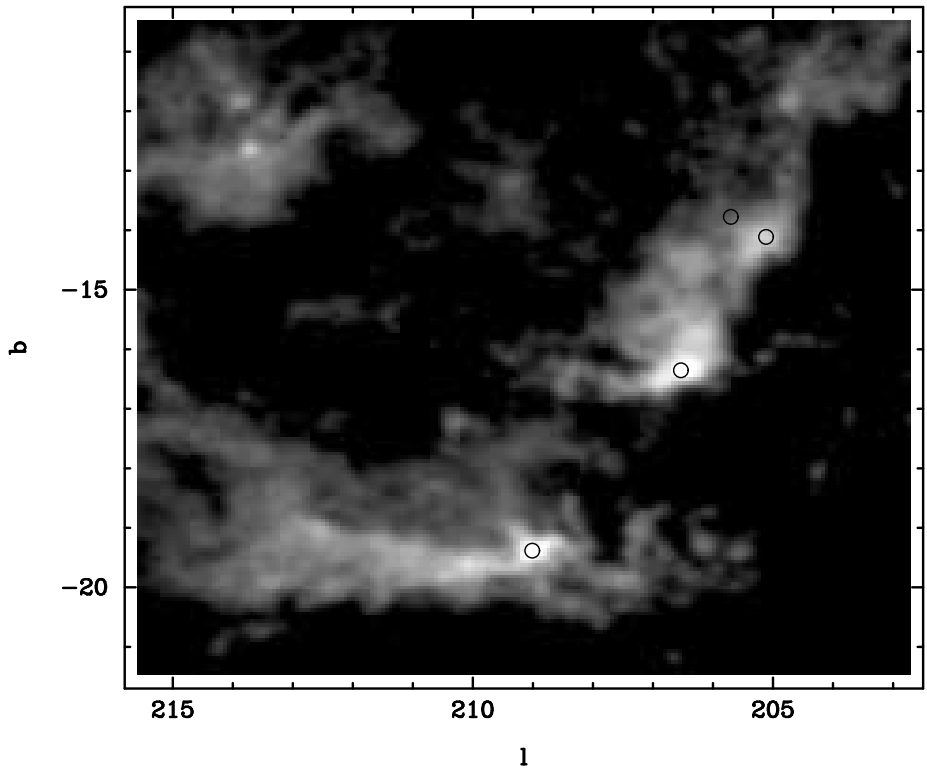}
\caption{Large-scale distribution of integrated $^{12}$CO emission in the region
of Orion, shown as a function of galactic longitude and latitude.
Positions of the Orion Nebula cluster and
the young NGC 2024, 2068, and 2071 clusters are marked by superimposed circles
(see text).  From Wilson (2001).}
\label{fig:orionco}
\end{figure}

Figure \ref{fig:orionco} shows the large-scale distribution of $^{12}$CO emission in the Orion
A and B clouds (Wilson 2001).  The overall morphology of the clouds suggests part
of an arc, such as might be produced by an expanding, flow-driven bubble
which accumulates material far out of the galactic plane (HBB01).
The overall structure is highly filamentary, especially in the A cloud.
Strikingly, the massive Orion Nebula Cluster (Hillenbrand 1997, and references
therein) and the young, dense embedded clusters NGC 2024, 2068, and 2071
(Lada 1992) lie preferentially at the ends of the molecular gas distribution,
just as would be predicted by the simplest version of sheet collapse in a
non-circular sheet.  There are multiple condensations of molecular gas
and young stars in these clouds, not just one major cluster at each end of each
cloud, but such independent condensations would occur as long as the
initial cloud were not a perfectly smooth ellipse in shape.

Dense clusters and dense filamentary gas lie only at one end of the Orion
A cloud (e.g., Ali \& Depoy 1995; Goldsmith, Bergin, \& Lis 1997).
The other (southern; higher galactic longitude $l$) 
end of the cloud appears to be much more diffuse and
contains only small groups of stars (e.g., Strom, Margulis, \& Strom 1989;
Strom, Strom, \& Merrill 1993).
We speculate that this difference is due to initial conditions; the
cloud prior to collapse was initially much more diffuse at one end than
the other.  The overall structure of the A cloud suggests a ``V'' shape,
with the dense narrowest region at the northern end (the region of the
so-called ``integral-shaped'' filament; see Bally \etal 1987).
Now, prior to overall collapse to a filament, our calculations for initially
elliptical sheets show similar structure at each end; denser concentrations
are formed at the ``tip'' of the ellipse, with two ``filaments'' streaming
out on either side.  We speculate that we are seeing a similar effect in the
A cloud; the southernmost parts have not collapsed as far as the northern
(lower -$l$) region.

\begin{figure}
\includegraphics[width=4in,angle=-90]{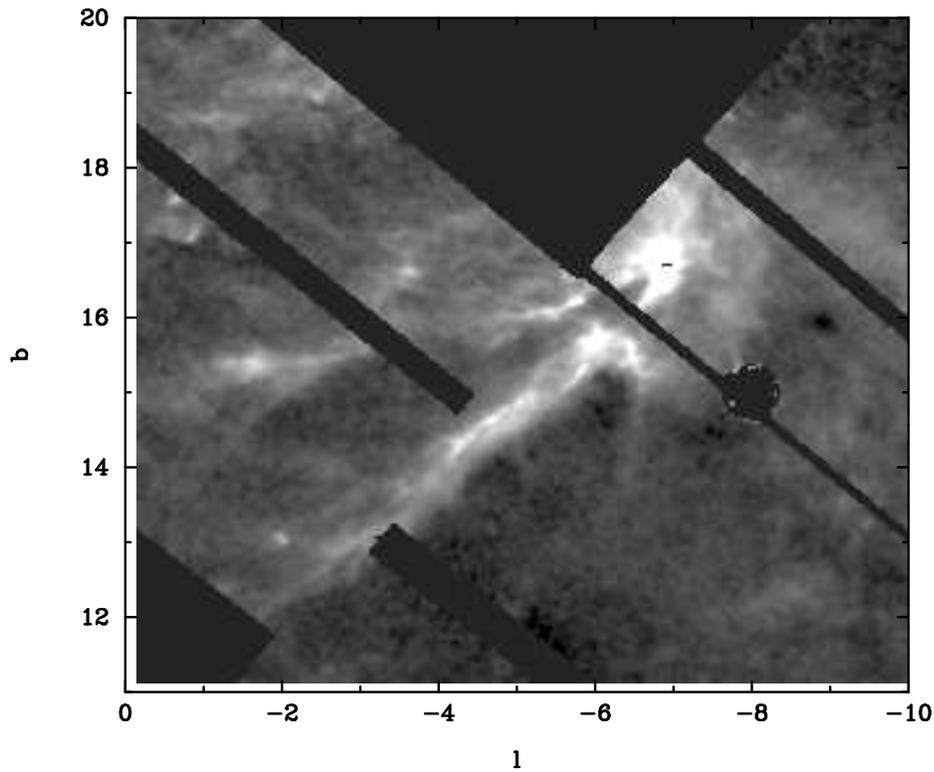}
\caption{ Extinction map of the Ophiuchus region, made by the COMPLETE project
using 2MASS data (Goodman 2004: see also http://cfa-www.harvard.edu/COMPLETE).
The major filamentary structures of the cloud and the main concentration of dust
(and gas) are evident (see text)}
\label{fig:ophext}
\end{figure}

Figure \ref{fig:ophext} shows an extinction map of the Ophiuchus region, which more or
less indicates the large-scale morphology of the molecular gas.  The well-known
filamentary structure extending outward from the main concentration of gas and
dust is evident.  Again, we speculate that the overall structure of this region
is due to a collapse similar to that shown at either end of our elliptical sheet
calculations, with a ``V'' of filaments extending out from the main dense
collapse region.  The structure is more complex than that of our elliptical
sheet simulations, but then the initial conditions are unlikely to be as smooth
and simple for real clouds.

\begin{figure}
\epsscale{0.5}
\plotone{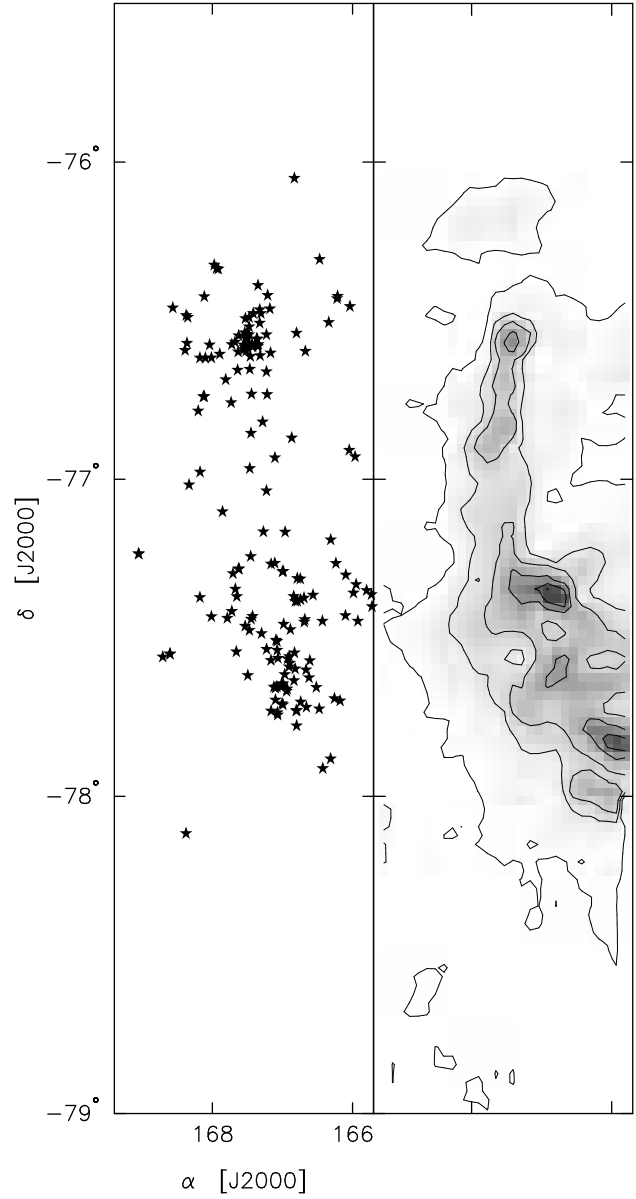}
\caption{Positions of young stellar members (left panel) and extinction contours
(right panel) for the Cha I star-forming region, modified from Carpenter \etal (2002)
(see text)}
\label{fig:cha}
\end{figure}

Figure \ref{fig:cha} shows the distribution of young stars and extinction (which, again, traces
the molecular gas fairly well) in the Cha I cloud.  Note how the cloud is filamentary, and
that once gain there are two clear concentrations of stars nearer the ends of the cloud.

\begin{figure}
\includegraphics[width=4in,angle=-90]{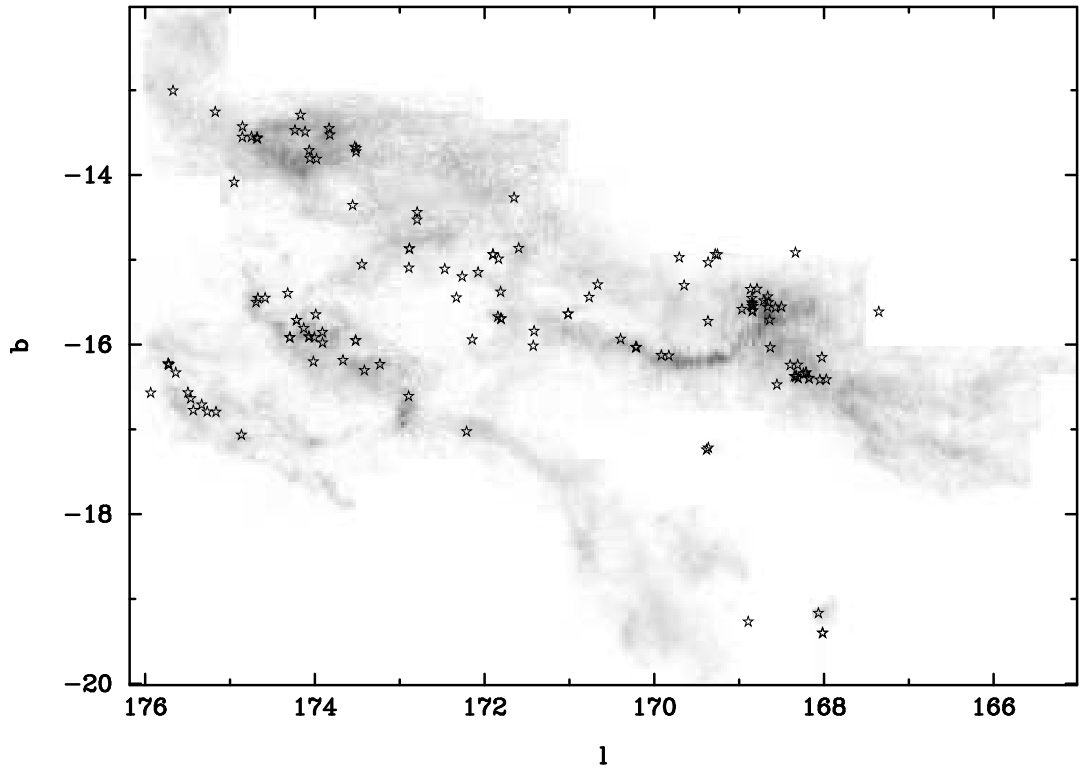}
\caption{Positions of young stars and protostars superimposed upon the
$^{13}$CO integrated emission in Taurus, the latter taken from Mizuno \etal;
(1995) (see text)}
\label{fig:tau}
\end{figure}

Not all molecular clouds exhibit a simple global filament structure with clusters at the endS.
Figure \ref{fig:tau}  shows the positions of the young stars in the Taurus
molecular cloud, superimposed on the $^{13}$CO integrated emission
(Mizuno \etal 1995).  As noted before (e.g., Hartmann 2002, and references
therein), Taurus is composed of extended, roughly parallel bands or
filaments of both gas and stars; gravitational fragmentation into several filaments
may have occurred (Miyama \etal 1987a,b; Nakajima \& Hanawa 1996). 
There are no major clusters of stars
in Taurus (although there is a small double group of stars in L1495; see
Figure (\ref{fig:tau}). 
Taurus is one of the most dispersed, extended, and low-density clouds, much more
extended than regions comparable in mass sich as Ophiuchus. We suggest that the
small-scale density and velocity fluctuations inevitably present in any realistic
scenario of cloud formation play a much larger role in Taurus than in other regions;
the low surface density suggests that global gravitational collapse may not dominate
the structure imposed by initial inhomogenities, in constrast with higher surface density
regions.

However, even in this case of Taurus there is some suggestive substructure. For
instance, the double group of L1495 seems to lie at the end of a filament and V-shaped
structure running from $l \sim 166$ to about $l \sim 169$; this gas structure seems 
distinct from other regions, especially as the radial velocities of the gas increase 
(to positive values) with increasing $l$, whereas the overall trend in Taurus is increasing
radial velocities with decreasing $l$ (e.g., Mizuno \etal 1995). Similarly, there is
structure near $l \sim 174$, $b \sim -13.5$ (Heiles Cloud 2) which exhibits a curious
oval shape with an interior hole as seen in integrated intensity; the young stars also
lie along the edge of the oval, suggesting fragmentation in a collapsing cloud edge.

It is worth noting that a number of molecular clouds show a rotation or
shear in the line of sight that is comparable to the gravitational
acceleration, such as the Orion complex (Bally \etal 1987) and Taurus
(e.g., Mizuno \etal 1997).  As shown in the simulation of Figure \ref{fig:rotateellipse}
such rotation can slow or prevent the overall collapse of the filament before fragmenting
and presumably forming stars.  A plausible scenario would be to assume sheets
with some angular momentum, insufficient to prevent collapse to a filament, but large
enough that the resulting filament does not collapse completely.  In this kind
of picture, there would be a tendency to form filaments with significant rotational support;
they would tend to shrink down until arriving at the angular momentum ``barrier''. 

In summary, we find that several local cloud complexes have morphologies
suggestive of the simplest versions of global collapse from a sheetlike configuration;
i.e., roughly filamentary cloud structure with concentrations of mass at the
end(s) of the clouds.

\subsection{Cluster formation}

The simple simulations discussed here may also have particularly important implications for the
formation and evolution of star clusters.  Many treatments of young clusters assume something
like an initial equilibrium configuration and follow the subsequent evolution.
However, the simulations presented here suggest that the accumulation of protocluster gas
is often a result of gravitational focussing; in other words, that the gas forming
the stars is initially collapsing.  Formation in collapsing media might result in violent
relaxation determining the cluster structure rather than two-body interactions, a result
suggested for the very young Orion Nebula Cluster by Hillenbrand \& Hartmann (1998).
Violent relaxation is not restricted to initially highly gravitationally unstable
conditions, like a collapsing sheet. In the absence of periodic boundary conditions, initially 
stabilized but efficiently dissipating turbulent clouds will evolve into global gravitational collapse
while fragmenting with violent relaxation also playing some role in the late phases of evolution
(Bate, Bonnell \& Bromm 2003; Bonnell, Bate \& Vine 2003).

Additionally, the picture presented here of cluster formation is
consistent with the ideas of competitive accretion forming massive stars
at the bottoms of cluster gravitational potential wells (Zinnecker 1982; Bonnell et. al. 2001a,b). 
It is worth noting that 
global infall into focal points can result in very high local mass infall rates, such as are needed to form
very massive stars in short times.  Alternative pictures in which high infall rates
are achieved in static clouds of order one Jeans mass by invoking a high turbulent velocity
to support the required high densities seem implausible. 
In our picture, global collapse could constitute a substantial fraction
of the observed ``turbulence'' in dense cores, with perhaps smaller-scale structure
generated by attraction to local mass concentrations. 

\subsection{Kinematics}

A further implication of the simulations is that the ``turbulent'' motions of many
star-forming structures are not necessarily those of a Kolmogorov spectrum,
but those of gravitationally-induced flows with substantially more power on large
scales.  Another way of saying this is that a substantial component of the observed
supersonic line widths in star-forming regions could be the result of collapse rather than
small-scale, random turbulent motions.  Our simulations are not ideal for exploring
this possibility; by restricting the motion to two dimensions and limiting the spatial
resolution, we are unable to follow details of the motion.  Nevertheless, the idea
of global collapse as an important generator of supersonic ``turbulence'' is very
attractive, in that some mechanism must be invoked to make gas concentrations in
the first place, and star formation must involve gravitationally-bound entities.

It is worth noting that, while our non-rotating and non-expanding simulations result ultimately
in collapse of all the material to the origin, many real clouds exhibit large scale velocity
gradients along their lengths (e.g., Bally \etal 1987), of a magnitude comparable to
that required to prevent total collapse.  Such velocity gradients must be the result of
initial conditions which generally provide molecular clouds with significant angular momenta.

\subsection{Implications for the initial mass function}

Our results suggest that there might be some relation between the boundary structure of
molecular clouds and the mass distributions of gravitationally-focussed concentrations, i.e.
between cloud edges and stellar/cluster mass functions.  Larson (1992) suggested that
fractal structure in clouds might be related to the stellar initial mass function; he speculated
that the observational indication of fractal projected cloud boundaries (e.g., Falgarone \etal 1991)
with fractal dimension $D \sim 1.35$ could be translated into a mass function $dN/d \log M \propto M^{-x}$,
with $x \sim 2.35$, consistent with the upper end of the stellar mass function (see also
Elmegreen 1997).  Our simulations suggest a physical mechanism -- gravitational focussing --
which can act {\em directly} on cloud boundaries to form mass concentrations,
with a distribution that reflects the size distribution of irregularities at
cloud boundaries.  This idea needs further exploration, as subsequent fragmentation and/or
competitive accretion could easily modify the mass function resulting simply from edge collapse.

\section{Conclusions}

Using numerical simulations of simple, isothermal, finite sheets, we have shown that gravity acting
on sheet edges can produce a wide variety of structures which are likely to have some relevance
to observed star-forming structures in molecular clouds.  In particular, we have shown that a likely
general result of the collapse of a sheet formed by flows in the ISM is a filament with higher
mass concentrations at the ends of the filament.  Any departure from circular symmetry at the
edge of gravitationally-bound clouds will tend to produce denser concentrations that may be
the origin of star clusters. Already in 1983 Bastien found that 
elongated cylindrical clouds fragmented into two condensations which he 
identified as an 'end effect' which
results from a similar physical behaviour as collapsing finite sheets.
We have shown that several nearby clouds exhibit morphologies
which are broadly consistent with the simulations.

We have addressed the problem of finite self-gravitating sheets 
in as simple a form as possible, limiting the motion to two dimensions.  
Even with these restrictions, our results emphasize the long-range effects of
gravity, and the importance of cloud boundaries, in generating structure and
turbulence.  
It is likely that clouds are formed with much more structure than assumed here;
further steps needed include simulating the formation of molecular clouds from
the diffuse interstellar medium, to explore what initial density and velocity
fluctuations are present.  The dynamic nature of even the simple simulations
presented here makes it likely that quasi-equilibrium treatments
of molecular cloud structure and star formation are unlikely to be realistic.
Our initial explorations emphasize the importance of gravitational focussing in
creating structure and turbulence in (finite) molecular clouds, a viewpoint that may lead
to new observational and theoretical approaches to understanding star formation.

\acknowledgments

Thanks to John Carpenter for providing a modified version of the Cha I cloud results and
to Alyssa Goodman for the extinction map of Ophiuchus.  This work was supported in part by
NASA grant NAG5-9670 and NAG5-13210.

\end{document}